\documentclass[a4paper,aps,jcp,floatfix,unsortedaddress,superscriptaddress,reprint]{revtex4-1}
\usepackage{amssymb,amsfonts,amsthm,bm}
\usepackage[utf8]{inputenc}

\usepackage{amsmath,bm}
\usepackage{graphicx}
\usepackage{color}
\usepackage{verbatim}
\usepackage[position=t,singlelinecheck=false,caption=false]{subfig}

\newcommand{\celsius}{\,^{\circ}{\rm C}}
\newcommand{\angstrom}{\textup{\AA}}

\newcommand*{\plimsoll}{{\ensuremath{-\kern-4pt{\ominus}\kern-4pt-}}}

\graphicspath{{figures/}}
\begin{document}

\title{Modelling toehold-mediated RNA strand displacement}

\author{Petr \v{S}ulc}
\affiliation{Center for Studies in Physics and Biology, The Rockefeller University, 1230 York Avenue, New York, NY 10065, USA}
\affiliation{Rudolf Peierls Centre for Theoretical Physics, 
 University of Oxford, 1 Keble Road, Oxford, OX1 3NP, United Kingdom}

\author{Thomas E.~Ouldridge}
\affiliation{Rudolf Peierls Centre for Theoretical Physics, 
 University of Oxford, 1 Keble Road, Oxford, OX1 3NP, United Kingdom}

\author{Flavio Romano}
\affiliation{Physical and Theoretical Chemistry Laboratory, 
 Department of Chemistry, University of Oxford, South Parks Road, 
 Oxford, OX1 3QZ, United Kingdom}

\author{Jonathan~P.~K. Doye}
\affiliation{Physical and Theoretical Chemistry Laboratory, 
 Department of Chemistry, University of Oxford, South Parks Road, 
 Oxford, OX1 3QZ, United Kingdom}

\author{Ard~A. Louis}
\affiliation{Rudolf Peierls Centre for Theoretical Physics, 
 University of Oxford, 1 Keble Road, Oxford, OX1 3NP, United Kingdom}

%
%


\begin{abstract}
{We study the thermodynamics and kinetics of an RNA toehold-mediated strand displacement reaction with a recently developed coarse-grained model of RNA. Strand displacement, during which a single strand displaces a different strand previously bound to a complementary substrate strand, is an essential mechanism in active nucleic acid nanotechnology and has also been hypothesized to occur {\it
 in vivo}. We study the rate of displacement reactions as a function of the length of the toehold and temperature and make two experimentally testable predictions: that the 
displacement is faster if the toehold is placed at the $5^{\prime}$ end of the substrate and that the displacement slows down with increasing temperature for longer toeholds.}
\end{abstract}

\maketitle

\section*{INTRODUCTION}
The emerging field of RNA nanotechnology aims to construct nanoscale structures and devices by using RNA strands \cite{guo2010,grabow2014}. It is closely related to the more established and rapidly developing field of DNA nanotechnology, which exploits the  specificity of Watson-Crick base pairing to construct impressive artificial nanoscale structures and active devices \cite{Seeman1982,Rothemund06,Douglas09,Douglas12,castro2011,linko2013enabled,Yurke2000,yurke2003using}.
While RNA and DNA are similar molecules, composed of a sugar-phosphate backbone to which an alphabet of four different nucleobases  can attach, there are some differences. For example in DNA the sugar is deoxyribose and the four bases are (A, C, G, T)  whereas in RNA  the sugar is ribose, and the base T is replaced by U. Like DNA, RNA can form two kinds of Watson-Crick base pairs (AU or GC), but it has a higher propensity for other kinds of bonding, including the wobble base pair (GU) as well as numerous other tertiary structure interactions. 

 In the cell, both molecules can store information, but whereas isolated DNA is primarily found in a B-helical double-stranded state, RNA is more versatile. It can fold from single-stranded states into complex three-dimensional structures that contain A-form helical double-stranded segments as well as loops, bulges and junctions. This increased structural repertoire facilitates biological functionality, so that RNA molecules can perform multiple additional roles, including catalysis, genetic regulation, structural support, and templating for molecular recognition and DNA synthesis \cite{Elliott11,cell,cech1986biological}. Since it can accomplish both storage of genetic material (like DNA) as well as metabolism (like proteins), it has been postulated that early life was based on RNA before DNA-based organisms appeared (the ``RNA world'' hypothesis \cite{gilbert1986origin}).  On the one hand, this versatility makes the prospect of using RNA nanotechnology very appealing, especially in biomedical applications.  On the other hand, it also makes predicting the three-dimensional structure significantly more challenging \cite{guo2010,laing2010computational,laing2011computational} than for DNA. For example, rather than designing {\it de novo} sequences that would fold into a particular functional three-dimensional structure, RNA nanotechnology often relies on functional motifs from known biologically occurring structures~\cite{grabow2013rna}.

In this paper we study the RNA equivalent of toehold-mediated strand displacement, a relatively simple dynamic reaction that has been a key component in many active devices in DNA nanotechnology. During strand displacement, a single-stranded ``invading'' strand replaces an ``incumbent'' strand that was bound in a duplex with a ``substrate'' strand. Both the invading and incumbent strand are complementary to the substrate strand, but the incumbent strand is a few bases shorter. When bound to the incumbent strand, the substrate strand hence has a short single-stranded overhanging region (a ``toehold'') to which in invading strand can bind.

Systems based on DNA strand-displacement reactions have been shown to be able to perform computation \cite{soloveichik2010dna,Qian2011,Seelig2006} and are the basis of autonomous DNA motors \cite{Muscat2011,Wickham12}. Strand-displacement reactions also have a great potential for use in RNA nanotechnology applications. For instance, a series of different reactions that involve several RNA strand-displacement steps were developed in Ref.~\onlinecite{lisapeng}. The strand displacement reaction is triggered by the presence of an mRNA strand with a particular sequence and the final product is an siRNA complex \cite{lisapeng}. One promising application {\it in vivo} is the conditional knockout of a gene by an RNA silencing mechanism in the presence of mRNA created by a transcription of the triggering gene. It was recently shown {\it in vitro} that a cascade of RNA strand displacement reactions can be used for the detection of a single-stranded RNA with a particular sequence that triggers the reaction \cite{sternberg2014exquisite}. Furthermore, RNA switches have been recently introduced for use in synthetic gene circuits, where the presence of a specific trigger RNA strand opens an RNA hairpin via a stand-displacement reaction. The hairpin loop contains a ribosome binding region that becomes accessible for transcription after the successful completion of the displacement. The functionality of RNA switches has been demonstrated {\it in vivo} as well as for diagnostics {\it in vitro} \cite{green2014toehold,pardee2014paper}. 

 There is currently substantial evidence that there are multiple regions of the genome that are transcribed into RNA molecules which are not further involved in protein production, but are themselves the final product. Thousands of these ``non-coding RNAs'' (ncRNA) have been identified and their function is a very active field of research \cite{washietl2007structured,djebali2012landscape,guttman2012modular}.
 Some of the ncRNA interactions also involve RNA-RNA contacts and the creation of RNA double-stranded sections \cite{hao2011quantifying,gottesman2002stealth}. It is hence plausible that RNA strand displacement might be involved in ncRNA interactions. 
 For instance, it was hypothesized in Ref.~\onlinecite{homann1996dissociation} that RNA strand displacement reactions can occur {\it in vivo} for example in ribozyme-product complexes and it was shown {\it in vitro} that a single-stranded RNA can displace the cleaved strand which was bound to a ribozyme \cite{homann1996dissociation}. More generally, the biological versatility of RNA makes it a prime candidate for many different {\it in vivo} applications of nanotechnology \cite{takahashi2013modular,guo2010,grabow2013rna}.  

 In this work we apply a recently developed coarse-grained model of RNA, oxRNA \cite{oxRNA}, to study the biophysics of toehold-mediated RNA strand displacement reactions. 
 OxRNA's development followed the coarse-graining approach used in a previously developed coarse-grained model of DNA, oxDNA \cite{Ouldridge2011,Ouldridge_tweezers_2010,Ouldridge_thesis,necitujoxDNA}, which was shown to accurately reproduce the kinetics and thermodynamics of DNA strand displacement \cite{srinivas2013biophysics}, including phenomena such as the effect of mismatches on displacement rates \cite{machinek}. 

 While the thermodynamics and kinetics of DNA strand displacement has been previously carefully studied both by experiments and by simulations \cite{Zhang2009,srinivas2013biophysics}, no systematic study is available for RNA. However,
 if the RNA strand displacement-based devices are to realize their full potential, an understanding of the underlying mechanisms is necessary. While one might expect the general features of the reaction to be similar that for DNA, we will also demonstrate interesting differences.
 
 This paper is organized as follows.
 We introduce the RNA model and the simulation methods in the next section. 
  We then study the free-energy profile of the reaction, the rates of displacement as a function of the length and the placement (at either the $3^{\prime}$ or $5^{\prime}$ end) of the toehold, and the effect of temperature on the rates. We find that the reaction rate is increased by six orders of magnitude when the toehold length increases from 1 to 6 bases at $37\celsius$. In contrast to DNA behavior, we find a noticeable difference in the displacement rate depending on whether the toehold is placed at the $3^{\prime}$ or $5^{\prime}$ end, with the $5^{\prime}$ being faster. Finally, we observe that the rate of displacement decreases with increasing temperature for a three-base toehold, but by less than expected from simply the decrease in stability of the toehold.
  

\section*{MODEL AND METHODS}
\subsection*{A coarse-grained model of RNA}
\label{sec_model}
We give a brief description of the oxRNA model here, while the details of the structural, mechanical and thermodynamical properties of the model are provided in Ref.~\onlinecite{oxRNA}. 

\begin{figure}[t]
\includegraphics[width=0.55\textwidth]{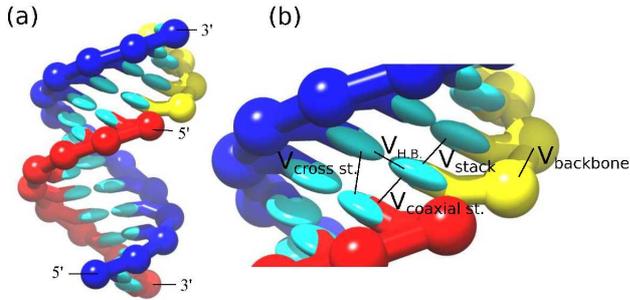}
\centering
\caption{A schematic representation of (a) an A-RNA helix as represented by the model and of (b) the attractive interactions in oxRNA \cite{oxRNA}.}
\label{fig_interactions}
\end{figure}

OxRNA represents each nucleotide as a single rigid body with multiple interaction sites. The rigid bodies interact with effective anisotropic interactions that are designed to
capture the overall thermodynamic and structural consequence of the base-pairing, stacking and backbone interactions, as schematically shown in Fig.~\ref{fig_interactions}.
The potential function of the oxRNA model is 
\begin{eqnarray}
\label{eq_potential}
 V_{\rm oxRNA} &=& \sum_{\left\langle ij \right\rangle} \left( V_{\rm{backbone}} + V_{\rm{stack}} +
V^{'}_{\rm{exc}} \right) +  \\
   &+& \sum_{i,j \notin {\left\langle ij \right\rangle}} \left( V_{\rm{H.B.}} +  V_{\rm{cross~st.}}  + V_{\rm{exc}}  + V_{\rm{coaxial~st.}} \right), \nonumber 
\end{eqnarray}
where the first sum runs over all pairs of nucleotides which are nearest neighbors on the same strand and the second sum runs over all other pairs. 

The backbone interaction, $V_{\rm{backbone}}$, is an isotropic FENE (finitely-extensible nonlinear elastic) potential
and depends only on the distance
between the backbone sites of the two adjacent nucleotides. This potential is
used to mimic the covalent bonds in the RNA backbone that constrain the intramolecular distance between neighboring nucleotides. The nucleotides also
have repulsive excluded-volume interactions $V_{\rm{exc}}$ and $V^{'}_{\rm{exc}}$ that depend on the distance between their interaction
sites, namely the backbone-backbone, stacking-stacking and stacking-backbone distances.
The excluded-volume interactions ensure that strands cannot overlap, or pass through each
other in a dynamical simulation.

The duplex is stabilized by hydrogen bonding ($V_{\rm{H.B.}}$), stacking
($V_{\rm{stack}}$) and  cross-stacking ($V_{\rm{cross~st.}}$) interactions.
These potentials are highly anisotropic and depend on the distance between the relevant
interaction sites as well as the mutual orientations of the nucleotides.  

The hydrogen-bonding term $V_{\rm{H.B.}}$ is designed to capture the
duplex-stabilizing interactions between Watson-Crick and wobble base pairs.
The stacking interaction $V_{\rm{stack}}$ mimics the favorable interaction between adjacent bases, which results from a combination of hydrophobic, electrostatic and dispersion effects.

The cross-stacking potential, $V_{\rm{cross~st.}}$, is designed to capture the
interactions between diagonally opposite bases in a duplex and has its minimum when the distance and mutual orientation
between nucleotides corresponds to that for a nucleotide and the
3$^\prime$ neighbor of the directly opposite nucleotide in an A-form helix. This interaction has been parametrized to capture the stabilization of an RNA duplex by a 3$^{\prime}$ overhang \cite{mathews1999expanded}. OxRNA does not include any
interaction with the
5$^\prime$ neighbor of the directly opposite nucleotide, as 5$^\prime$ overhangs are significantly less
stabilizing than 3$^\prime$
overhangs \cite{mathews1999expanded}.  

The coaxial stacking potential $V_{\rm{coaxial~st.}}$  represents the
stacking interaction between nucleotides that are not nearest neighbors on the
same strand.

The model currently does not include explicit electrostatic interactions. It was parametrized to reproduce RNA behavior at high ($1\,\rm{M}$) salt concentrations where the long-range interactions between phosphate charges on the backbone are screened. The remaining short-range electrostatics repulsion interactions are incorporated into the excluded volume potentials. 

The interactions in the model were parametrized to reproduce the melting temperatures of hairpins and short oligomers as predicted by the nearest-neighbor model of RNA thermodynamics \cite{mathews1999expanded,xia1998thermodynamic} that is the basis of most RNA secondary structure prediction tools \cite{nupack,kinefold,ViennaRNA}. Two parametrizations of oxRNA are available, the ``sequence-averaged'' and the ``sequence-dependent'' version. 
In the sequence-dependent parametrization, the hydrogen bonding and stacking interactions have different strengths depending on the types of the interacting bases. The sequence-averaged interaction strengths do not distinguish between the different base types, and in particular the interaction strengths of the hydrogen-bonding potentials are the same for Watson-Crick base pairs (AU and CG) and zero otherwise. The sequence-averaged interaction strengths were fitted to reproduce the thermodynamics of an averaged nearest-neighbor model, where the respective free-energy contributions were averaged over all possible base-pair steps with Watson-Crick complementary base pairs. 

In this work, we use the sequence-averaged parametrization, as we aim to study the displacement process for different toehold lengths and temperatures without the complications of sequence-specific effects. Our results will be comparable to experiments where ``average'' sequences are used, i.e.~sequences with similar stability as the ones obtained from the averaged nearest-neighbor model for a given sequence length. 

\subsection*{Simulation methods}
\label{sec_methods}
The thermodynamics of the strand-displacement process is probed using the virtual move Mote Carlo (VMMC) algorithm (specifically the variant described in the Appendix of Ref.~\onlinecite{Whitelam2009}) combined with umbrella sampling to help the system overcome free-energy barriers \cite{Torrie1977}. The chosen order parameters were the number of bonds between the invader and the substrate strand and bonds between the incumbent strand and the substrate. The weights assigned to the respective states were chosen by experience and then adapted by hand to ensure an efficient sampling.

We use the forward flux sampling (FFS) method \cite{Allen2009} to estimate the rate of the strand displacement reaction.
The FFS technique facilitates simulation of transition pathways by splitting up the rare transition event into multiple stages which are sampled separately.
We previously used FFS simulations with the coarse-grained model of DNA, oxDNA, to study the kinetics of DNA strand displacement \cite{srinivas2013biophysics} and hybridization \cite{ouldridge2013dna,schreck2014dna}. In this work, we use Direct FFS (described in detail in the Supplementary Material of Ref.~\onlinecite{ouldridge2013dna}).
The FFS simulations are run after an initial equilibration of $10^5$ steps. The initial flux is measured through the interface at which the minimum distance between the complementary bases of the invading strand and the substrate strand is smaller than $0.84\,\angstrom$. We then measured the probability of successfully crossing successive interfaces that were defined by the number of the formed bonds between the invading strand and the substrate (see Supplementary Material for details). 

 In our simulations, we define a base pair as being formed if the hydrogen-bonding energy ($V_{\rm H.B.}$ in Eq.~\ref{eq_potential}) between the two bases is more negative than $- 1.0\, k_{\rm B}T$ for $T=300\,{\rm K}$, which corresponds to about 18\% of typical hydrogen-bonding energies in the oxRNA model. We do not only allow any misbonds, i.e.~only the designed `correct' base pairs between the invading strand and the substrate (and the incumbent strand and the substrate) can form.
 
The molecular dynamics simulations for the FFS were carried out with an Andersen-like thermostat (described in the Appendix of Ref.~\onlinecite{Russo09}).   
All simulations are carried out with a simulation box size that corresponds to an equal strand concentration of $65.3\, \mu\rm{M}$ for the invading strand and substrate.  Even though nucleic acid nanotechnology experiments are usually done at lower strand concentrations ($\sim \rm{nM}$ or low $\rm{\mu M}$), we chose a higher concentration for our simulations in order not to spend 
too much time simply simulating the diffusion prior to the encounter and reaction of the strands. Such an approach allows us to efficiently extract the relative second-order rate constants
\cite{srinivas2013biophysics,ouldridge2013dna}. 




The simulation step $\delta t$ was chosen to be $0.005$ in simulation time units, corresponding to $1.53 \times 10^{-14}\,{\rm s}$.
We further used a high translational diffusion constant for an RNA nucleotide in our simulations, $D = 5.8 \times 10^{-7}$\,m$^2$s$^{-1}$, which corresponds to a diffusion constant of
$2.1 \times 10^{-8}$\,m$^2$s$^{-1}$ for a 14-mer, larger than the experimentally measured $D_{\rm exp} = 0.92 \times 10^{-10}$\,m$^{2}$s$^{-1}$ \cite{Lapham1997}.
We intentionally use a higher diffusion constant in order to speed up the diffusion of the strands in our simulations. Again, as the motion of strands is still diffusive over appreciable timescales, this should speed up the simulation without altering the basic physics of the reactions. The rotational diffusion coefficient was set to $D_{\rm rot} = 3D$ and each nucleotide was treated as a rigid body with a diagonal inertia tensor. 

One needs to be cautious in interpreting time units in
coarse-grained simulations. For instance, time scales of different processes might scale with different factors \cite{padding2006hydrodynamic}. Furthermore, we use an artificially high diffusion constant.
 Therefore, rather than trying to map our simulation results onto experimental rates, our emphasis is on computing the relative rates of similar processes (e.g.~strand displacement with different lengths for the toehold) which can then be compared with experimental data, should they become available. 


\section*{RESULTS AND DISCUSSION}

\begin{figure}
\includegraphics[width=0.4\textwidth]{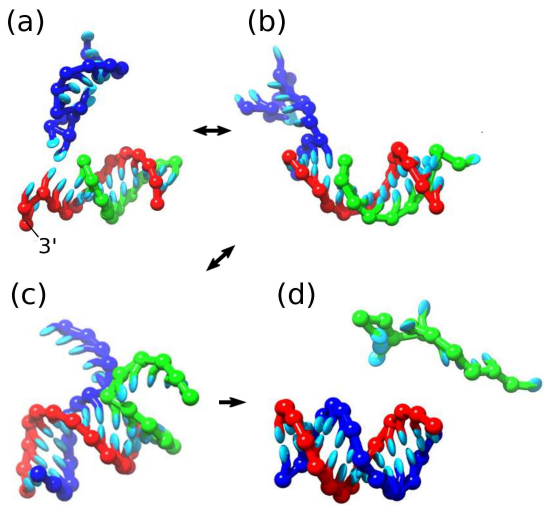}
\centering
\caption{A schematic representation of the different stages of a strand displacement reaction as represented by oxRNA for a four-nucleotide long toehold at the $3^{\prime}$ end of the substrate. (a) The invading strand (blue) attaches to the toehold of the substrate (red). (b) The invading strand is fully bound to the toehold. (c) The invading strand displaces bonds between the incumbent (green) strand and the substrate. (d) The incumbent strand loses all bonds with the substrate.}
\label{fig_stages}
\end{figure}


The different stages of the reaction are introduced in Fig.~\ref{fig_stages} as represented by the oxRNA model.
We note that the respective stages of the reaction are reversible, and it is possible for the invading strand to detach before successfully completing the displacement. 

The strand-displacement reaction in DNA has been studied in detail both experimentally \cite{yurke2003using,Zhang2009} as well as theoretically with the help of a simplified 1-D model, models based on secondary structure as well as a previously developed coarse-grained model of DNA, oxDNA \cite{yurke2003using,Zhang2009,srinivas2013biophysics}.

In Ref.~\onlinecite{Zhang2009}, the rate of the toehold-mediated strand displacement was measured as a function of the length of the toehold. For an average strength toehold (i.e. a sequence with roughly the same number of AT and GC bonds) it was found that the rate of the reaction increases exponentially with the length of the toehold until it saturates for a toehold of about six bases, where the rate is about $6.5$ orders of magnitude faster than for a system with no toehold. A very similar speed-up was also observed in coarse-grained simulations with oxDNA \cite{srinivas2013biophysics}. 

We are confident that our oxRNA model is able to provide insight into the RNA strand displacement reaction, as it adopts a very similar coarse-graining approach to that developed for oxDNA, which was shown to sufficiently describe the necessary biophysical properties of DNA to capture both thermodynamic and kinetic aspects of the strand displacement reaction \cite{srinivas2013biophysics}.

In the rest of this work, we will study the strand displacement reaction with toeholds of length ranging from 1 to 6, placed at either end of the substrate. The substrate has 10 base pairs with the incumbent strand. 
We use the oxRNA model to obtain a free-energy profile for the strand-displacement reaction, and the rates of the RNA displacement as a function of toehold length, position ($3^{\prime}$ or $5^{\prime}$) and temperature.

\subsection*{Free-energy profile}
\label{sec_profile}

\begin{figure}
\includegraphics[width=0.44\textwidth]{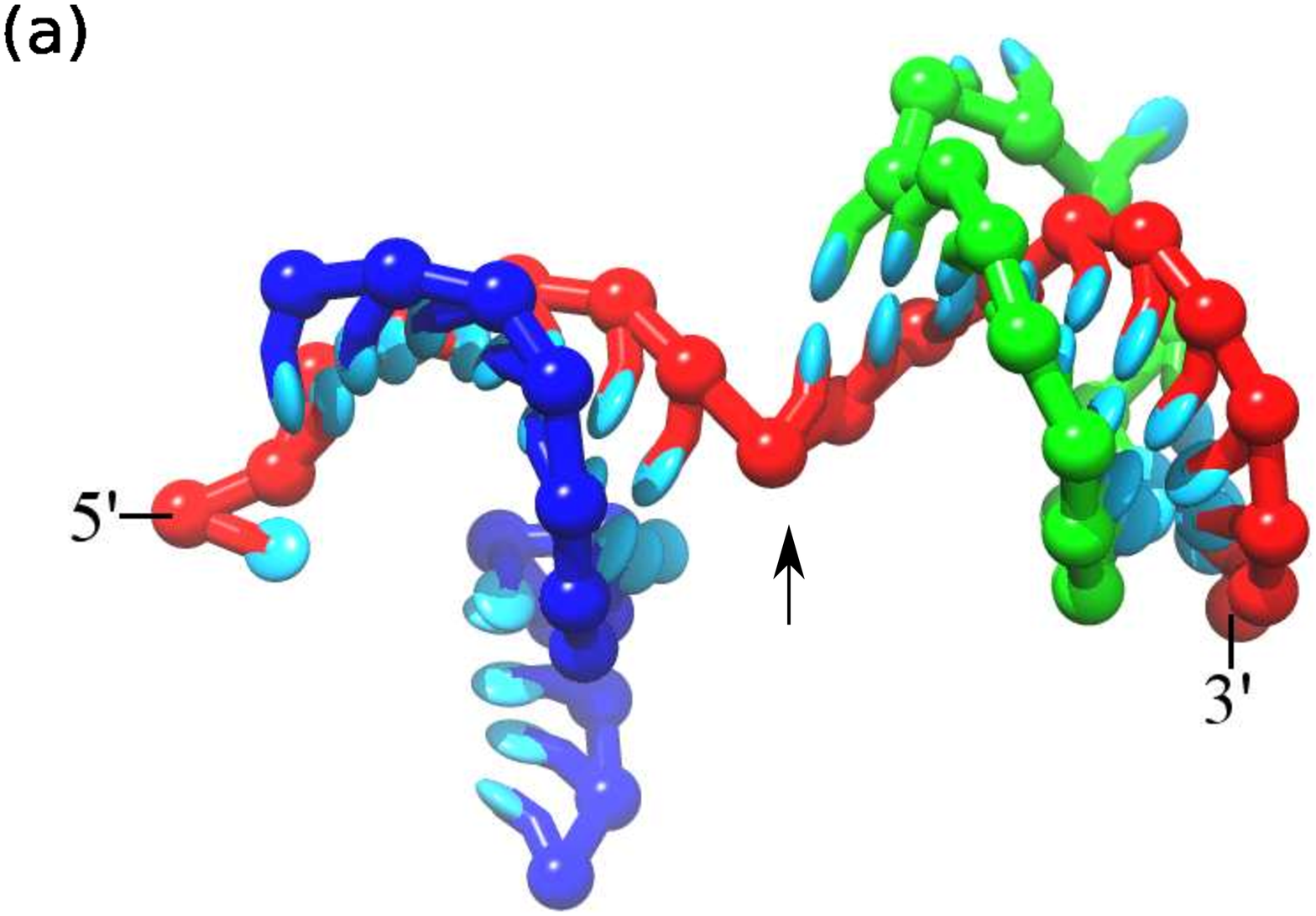}\\
\includegraphics[width=0.5\textwidth]{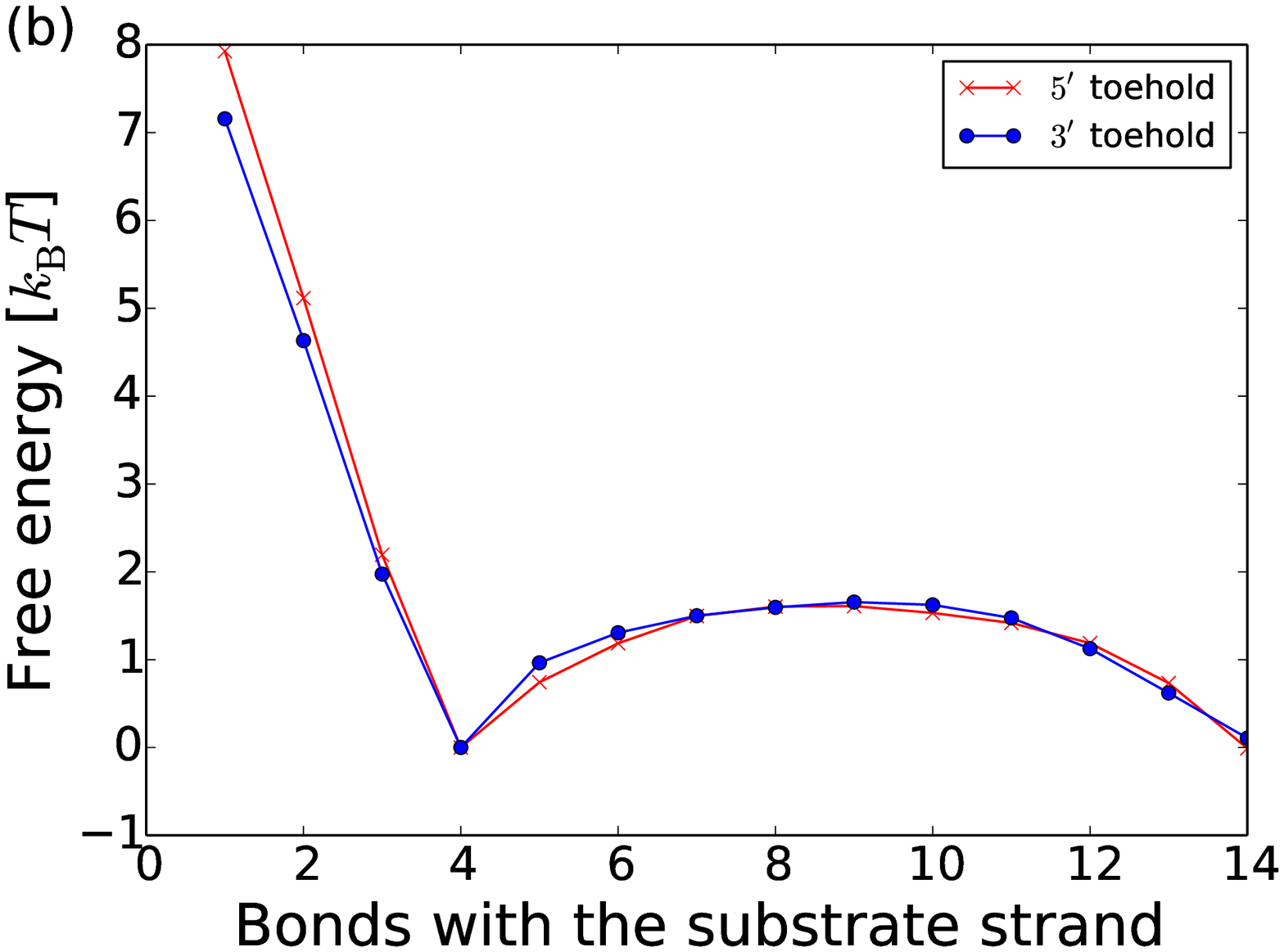}
\centering
\caption{(a) An 18-base substrate strand (red) with two 14-base strands attached at its $3^{\prime}$ and $5^{\prime}$ 4-base toeholds, respectively. The stacking interaction at the branch migration junction (indicated by the black arrow) is broken, making it easier for the single-stranded overhangs of the strands to avoid each other. (b) Free-energy profile as a function of the number of bonds between the invading strand and the substrate for a $3^{\prime}$ (blue dots) and $5^{\prime}$ (red crosses) toehold at $37\celsius$. The free energy has been set to 0 when the invading strand has 4 bonds with the substrate.}
\label{fig_free_energy}
\end{figure}

We measure the free-energy profile of the toehold-mediated strand displacement reaction at $37\celsius$ by sampling the states of a system consisting of a 18-nucleotide substrate strand with two 4-nucleotide toeholds at the $3^{\prime}$ and $5^{\prime}$ ends. Two 14-nucleotide strands (that displace each other) are attached to the substrate, one at each toehold. The system is illustrated in Fig.~\ref{fig_free_energy}(a). We use VMMC (combined with umbrella sampling) to sample the free-energy landscape as a function of number of the bonds between each strand and the substrate. We plot the free energy in Fig.~\ref{fig_free_energy}(b) as a function of the number of bonds between the invading strand and the substrate, separately for each invading strand. The VMMC algorithm was run for approximately $5\times 10^{11}$ cluster move attempts. The errors for each point, estimated as the maximum difference from the mean from 3 independent simulation sets, is at most $0.06\, k_{\rm B} T$, i.e.~not larger than the size of the points in Fig.~\ref{fig_free_energy}(b).
In the simulation, we only allowed base pairing between the substrate and its complementary bases on the other two strands. We required that the strands always have at least one bond with the substrate to prevent dissociation. 
%

The free energy decreases as more bonds are formed between the invading strand and the toehold and reaches its minimum when the invading strand is fully bound to the toehold.
There is a free-energy barrier associated with the initiation of the displacement (i.e. branch migration, where the invading strand gains additional bonds with the substrate at the expense of the incumbent strand) of about $1.6\,k_{\rm B} T$ (corresponding to about 1 kcal/mol), which saturates after about 4 bases have been displaced.
A similar barrier was observed for DNA displacement \cite{srinivas2013biophysics,necitujsulc2012simulating} and it arises from the fact that at the branch point (the point on the substrate where the invading and incumbent strands meet) the strands are in close proximity and the single-stranded overhangs of the invading and incumbent strands cannot overlap because of the excluded volume interactions. The two strands can move away from each other, but this requires the stacking interactions at the junction to be broken, as illustrated in Fig.~\ref{fig_free_energy}(a). Hence, there is a free-energy penalty associated with taking the first steps of branch migration, even though the total number of base pairs between the substrate and the invading and incumbent strands remains the same throughout the process.



When the invading strand is bound to the toehold, it will eventually either initiate branch migration or fall off. We note that binding to the $5^{\prime}$ toehold is more favorable than binding to the $3^{\prime}$ toehold, as the invading strand attached at the $5^{\prime}$ end has an additional cross-stacking interaction. 
Due to the A-form helical structure, the $3^{\prime}$ overhang cross-stacking interaction is stronger than the $5^{\prime}$ overhang, as the distance towards the $3^{\prime}$ overhang is smaller, hence allowing for a stronger interaction. The average extra stabilization at $37\celsius$ in the nearest-neighbor model of RNA thermodynamics \cite{mathews1999expanded} is about $-0.4\,k_{\rm B} T$ for a $5^{\prime}$ overhang and $-1.3\, k_{\rm B} T$ for a $3^{\prime}$ overhang. 

The cross-stacking interaction in the oxRNA model was parametrized to reproduce the average melting temperatures of short RNA duplexes with extra unpaired overhangs at the $3^{\prime}$ ends of the strand. The model does not include any stacking interaction with the $5^{\prime}$ overhang, as it is significantly weaker than for the $3^{\prime}$ overhang.

 

\begin{figure}
\includegraphics[width=0.4\textwidth]{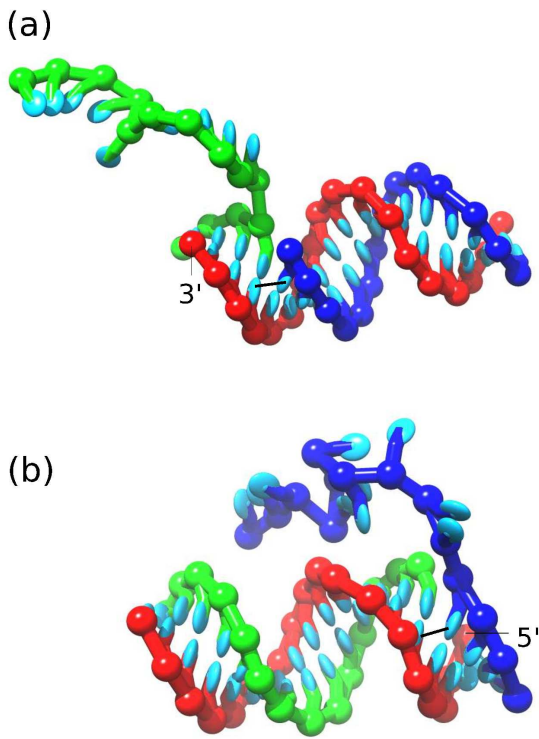}
\centering
\caption{(a) An invading strand (green) at a $3^{\prime}$ toehold has no cross-stacking interaction with the substrate (red) strand at the junction, as opposed to an invading strand (blue) attached to a $5^{\prime}$ toehold (shown in (b)). The $3^{\prime}$ cross-stacking interaction at the junction of the invading and incumbent strand is indicated schematically with a thick black line.}
\label{fig_3vs5end}
\end{figure}

The additional stabilization provided by the cross-stacking between the substrate and the invading strand attached to the $5^{\prime}$ toehold (Fig.~\ref{fig_3vs5end}(a)) is about $-1.7\,k_{\rm B} T$ in our model at $37\celsius$. Hence, the probability of falling off once the invading strand is attached is lower for an invading strand at a $5^{\prime}$ toehold than at a   $3^{\prime}$ toehold. 
%

While Fig.~\ref{fig_free_energy} provides information about the free-energy landscape of the strand displacement reaction as a function of the number of bonds between the invading strand and the substrate strand, it does not give us the rate of the reaction.
Even though some of the neighboring states during the branch migration procedure have almost no difference in free energy, the transition between them is a complicated process that requires the loss of a base-pair between the incumbent strand and the substrate and the creation of the base-pair between the invading strand and a substrate.

In a previous paper \cite{srinivas2013biophysics}, this
complex process was directly simulated for DNA with oxDNA. Furthermore, the transition rates between  adjacent states were described with a simplified intuitive  1-D model that included an effective free-energy barrier height that could be fit to reproduce the rates. However, these barriers cannot be obtained without either experimental or computational input.
To obtain kinetic information about the process, we hence use FFS simulations, as described in the following sections. 



\subsection*{Displacement rates for different toehold lengths and positions}
\label{sec_ratesA}
We calculate the relative rates of the strand displacement reactions using the FFS method, as outlined in the Methods section. 
We studied systems where the incumbent strand (shown in green in Fig.~\ref{fig_stages}) has 10 bonds with the substrate strand (shown in red in Fig.~\ref{fig_stages}). The substrate has a toehold of length $l$ that ranges from 1 to 6 bases. The invading (blue) strand is fully complementary to the substrate strand and can hence form up to $10 + l$ base pairs with the substrate strand once the incumbent strand is successfully removed. For each toehold length considered, we calculated the rates for toeholds placed at either the $3^{\prime}$ or $5^{\prime}$ ends of the substrate strand. All simulations were carried out at $37\celsius$.

\begin{figure}
\includegraphics[width=0.5\textwidth]{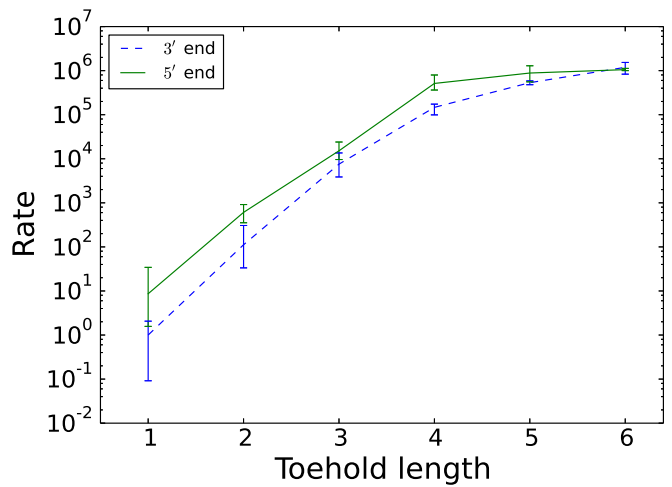}
\centering
\caption{Semilogarithmic plot of mean rates of the strand displacement reaction as a function of toehold length for toeholds at the $3^{\prime}$ end (dashed line) and at the $5^{\prime}$ end (full line) of the substrate strand. The errorbars show the maximum and minimum rate obtained from all the simulations for a given length of toehold. The rates are normalized with respect to the mean rate for a 1-nucleotide toehold at the $3^{\prime}$ end.}
\label{fig_rates_53}
\end{figure}

At experimental concentrations ($\mu{\rm M}$ or less), basic displacement reactions are well-described by second-order kinetics. For computational convenience, we simulate much higher concentrations of strands but this can potentially lead to the breakdown of the second-order description, due to reaction intermediates that are relatively long-lived compared to the overall reaction time scales. To infer the relative second-order rate constants at low concentration from our data, we neglect the time spent in these intermediate states once binding to the toehold has occurred in our calculation of the reaction rates from the simulations. Note that we do not preclude the possibility of displacement failure after binding to the toehold; we simply ignore the time that such a process takes. More details of this approach are provided in Refs.~\onlinecite{srinivas2013biophysics} and \onlinecite{ouldridge2013dna}.
 


For short toeholds, displacement will often fail and the invading strand undergoes frequent binding and unbinding to the toehold. For long toeholds the probability of falling off once fully bound to the toehold becomes very small and hence it will then successfully complete displacement with a high probability. Thus we expect the rate of the overall displacement reaction to increase with increasing toehold length up to a length at which it saturates.

The rates of displacement as a function of toehold length are shown in Fig.~\ref{fig_rates_53}. For each of the data points, we ran at least 3 independent sets of FFS simulations. 
We observe that the reaction rate of the toehold-mediated strand displacement reaction saturates for a toehold of around 5 nucleotides at a value that is approximately 6 orders of magnitude larger than the rate for a 1-nucleotide $5^{\prime}$ toehold and 5 orders of magnitude larger than for a 1-nucleotide $3^{\prime}$ toehold.

A similar range of relative rates was observed in DNA displacement reactions \cite{Zhang2009,srinivas2013biophysics}, where for the average toehold sequence, the rates saturated for a 6-nucleotide toehold at a value that is about $10^5$ times larger than the rate for a 1-nucleotide toehold. We note, however, that we studied RNA displacement kinetics at $37\celsius$, while the DNA displacement was studied at $25\celsius$. 
Furthermore, no strong dependence on whether the toehold was at the $3^{\prime}$ or $5^{\prime}$ end of the substrate was observed for DNA. We expect RNA to saturate for slightly shorter toeholds because an average RNA sequence with Watson-Crick base pairs is more stable than an average DNA sequence.


We note that until the toeholds are long enough to reach the saturated rate, the mean rates for the $5^{\prime}$ toeholds are larger than those for the $3^{\prime}$ toeholds. The ratios of the mean rates for different ends of the substrate are shown in Table~\ref{table_ratios}.
As discussed in the Free-energy profile section, the invading strand at the $5^{\prime}$ end gains additional stabilization from cross-stacking interaction with the substrate strand (as shown schematically in Fig.~\ref{fig_3vs5end}) and hence the probability for the invading strand to fall off from the $5^{\prime}$ toehold is lower and thus the displacement rate is increased. 

Our prediction of faster rates of displacement from a $5^{\prime}$ toehold is based on the experimentally verified fact that the cross-stacking with $3^{\prime}$ overhangs is significantly more stabilizing than the interaction with $5^{\prime}$ overhangs. The cross-stacking interaction in oxRNA captures the stabilization by a $3^{\prime}$ overhang, but the model does not include the stabilization by the $5^{\prime}$ overhangs, as they are significantly smaller. However, they still might contribute somewhat to the stabilization of the displacement from a $3^{\prime}$ toehold, which would mean that our inferred speed-up may be slightly overestimated. 


 \begin{table}
 \centering
 \begin{tabular}{ c c }
  \hline  \hline  {\bf Toehold length} & {\bf $5^{\prime}$ rate / $3^{\prime}$ rate} \\
  \hline
   1 & 8.6 \\ 
   2 & 5.4 \\ 
   3 & 2.0  \\
   4 & 3.5 \\
   5 & 1.7 \\
   6 & 0.9 \\ \hline \hline
 \end{tabular}
 \caption{The ratios of the mean displacement rates for toeholds at $5^{\prime}$ and $3^{\prime}$ ends as estimated from the FFS simulations at $37\celsius$.}
 \label{table_ratios}
 \end{table}

The rates were obtained with the averaged oxRNA model  
and hence apply to toeholds that would have stability similar to that of the averaged toehold (i.e.~with about the same number of AU and GC bonds). However, if one designed a system with a weak toehold (AU-rich), then the saturation would be reached at a toehold length higher than 5, and similarly, for strong (GC-rich) toehold, the saturation of the displacement speed would be reached at toehold lengths smaller than 5.  

 We further note that it is not only the number of AU and GC bonds that determines the free-energy stabilization for a particular sequence, but also their order \cite{oxRNA,necitujoxDNA}. Therefore, care needs to be taken in experiments when comparing for instance the rates
of the $3^{\prime}$ and $5^{\prime}$ toeholds, as the rates will be affected by the stabilization provided by the respective toehold sequences as well as the type of bases involved in stacking and cross-stacking interactions, which also varies with sequence \cite{mathews1999expanded}. 

Furthermore, we would expect the saturated rates to be different for weak and strong toehold sequences, with the strong ones being faster, as was previously observed for DNA \cite{Zhang2009}. 
It was observed for the oxDNA model \cite{ouldridge2013dna} that the hybridization rates are faster for stronger sequences, as the strand has higher probability of completely binding to the toehold region after the first few base pairs were made if the base pairs created during the first contact are stronger. We expect the same sequence-dependent effects to play a role in RNA strand displacement. However, given the absence of systematic rate data on RNA strand displacement to which to compare, we focus only on the average-model description in this work.

\subsection*{Displacement rate at different temperatures}
\label{sec_ratesB}

\begin{figure}
\includegraphics[width=0.5\textwidth]{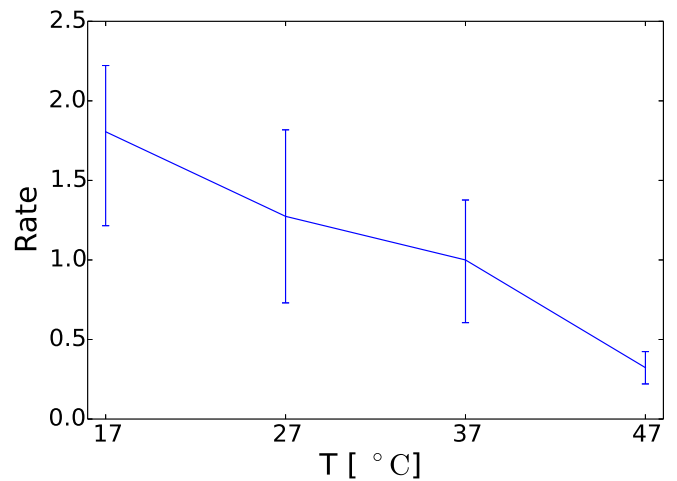}
\centering
\caption{Mean rates of the strand displacement reaction for a 3-nucleotide toehold at the $3^{\prime}$ end as a function of temperature. The errorbars show the maximum and minimum rate obtained from all the simulations for a given temperature. The rates are normalized with respect to the mean rate at $37\celsius$.}
\label{fig_rates_temp}
\end{figure}

We further investigate the relative displacement rates for a three-nucleotide toehold at the $3^{\prime}$ end for simulation temperatures ranging from $17$ to $47\celsius$.
For each temperature considered, we performed at least 3 independent sets of FFS simulations. The mean relative rates of the displacement are shown in Fig.~\ref{fig_rates_temp}. The rates decrease with increasing temperature, with the displacement at $47\celsius$ being $5.6$ times slower than at $17\celsius$. 

It is interesting to compare the temperature dependence of the rates with that for the yields of invading strands bound to the toehold only. We therefore also performed VMMC umbrella sampling simulations where the sampling of attachment and detachment of the invading strand to the three-nucleotide toehold at the $3^{\prime}$ end of the substrate was performed at $37\celsius$ and extrapolated to temperatures in the range $17$ to $47\celsius$ by a histogram reweighting method. In the simulation, the invading strand was allowed to make at most 4 bonds with the substrate strand. The VMMC simulations were run with the same simulation box size as the FFS simulations, corresponding to an equal strand concentration of $65.3\, \mu\rm{M}$. 
We observed that the decrease of the displacement rate with increasing temperature is smaller than the decrease of the yield of the invading strands bound to the toehold, as shown in Table \ref{table_temprates}.

 \begin{table}
 \centering
 \begin{tabular}{ c c c }
  \hline  \hline  {\bf T} & {\bf Displacement rate $k$} & $ k_{\rm on} / k_{\rm off} $  \\
  \hline
    $17 \celsius$ & $1.8$ &  $5.3$\\
   $27 \celsius$  & $1.3$  & $2.3$ \\ 
   $37 \celsius$  & $1$  & $1$ \\  
   $47 \celsius$  & $0.3$  & $0.4$  \\ \hline \hline 
 \end{tabular}
 \caption{The mean relative displacement rates $k$ (second column) compared with the normalized yield of invading strands attached to the toehold of the substrate (third column).  The rates and yields were normalized with respect to their respective values obtained at $37\celsius$. }
 \label{table_temprates}
 \end{table}

These results can be rationalized with a simple model for the toehold-mediated strand displacement reaction (see Ref.~\onlinecite{srinivas2013biophysics} or Supplementary Material S-I).
In the second-order limit that we consider (in which the lifetime of the three-stranded intermediate is neglected), the rate of displacement can be written as $k = k_{\rm on} \times p_{\rm bm | toe}$, where $k_{\rm on}$ is the rate of binding to the toehold and $p_{\rm bm | toe}$ is the probability of a successful completion of branch migration once bound by the toehold. Further, $p_{\rm bm | toe} = k_{\rm bm} /k_{\rm off}$, where $k_{\rm off}$ is the rate of unbinding from the toehold and $k_{\rm bm}$ is the rate at which the system initiates (and subsequently completes) branch migration. Thus  $k = k_{\rm on} \times k_{\rm bm}/k_{\rm off} $.



The quantity $k_{\rm on} / k_{\rm off}$ corresponds to the yield of invading strands attached to the toehold and is shown in the third column in Table \ref{table_temprates}. 
The difference in the temperature dependence of the displacement rate $k$ and the yield of toehold-bound duplexes
implies that $k_{\rm bm}$ must be affected
by the change of temperature as well. The implied increase in $k_{\rm bm}$ with temperature is presumably due to the disruption of base pairs involved in initiating branch migration and passage over the initial barrier to displacement highlighted in Fig.~\ref{fig_free_energy} is made easier by increased temperature.


As the toehold gets shorter, we would expect $k_{\rm on} / k_{\rm off}$ to become increasingly less dependent on temperature. Therefore, the temperature dependence of $k_{\rm bm}$ will be more dominant and we hence expect the slope of the graph of the displacement rate as a function of temperature to become less negative with decreasing toehold length.
Indeed, for the shortest toehold lengths (and certainly for blunt-ended displacement), we would expect the displacement rate to increase with temperature, as has been reported in DNA \cite{reynaldo2000kinetics}.




\section*{CONCLUSIONS}

We have studied toehold-mediated RNA strand displacement with a coarse-grained model of RNA, oxRNA \cite{oxRNA}. 
We observed behavior that was previously encountered in studies of DNA displacement \cite{srinivas2013biophysics}. In particular, we showed that there is a free-energy barrier to the initiation of the displacement process that is caused by the need for the single-stranded overhangs of the invading and incumbent strand to avoid each other (entropic effect) and that in doing so they break the stacking at the branch migration junction (enthalpic effect). We found the range of the relative speed-up of the displacement reaction (by about 5 to 6 orders of magnitude) between the saturated rate and the rate for a 1-base toehold for RNA is similar to that observed for DNA displacement.



We further found that for toeholds shorter than 5 nucleotides, i.e.~before the saturation regime is reached, the displacement reaction is about 2 to 9 times faster at the $5^{\prime}$ end than at the $3^{\prime}$ end. The invading strand gets additional stability from a cross-stacking interaction at the $5^{\prime}$ end toehold that reduces the probability of falling off from the toehold and hence increases the rate. Such an effect can be tested experimentally and can provide an additional way to modulate displacement rates.

Finally, we studied the kinetics of the displacement reaction for a three-nucleotide toehold at the $3^{\prime}$ end as a function temperature. The rates decrease by about a factor of 6 
when the temperatures is increased from $17$ to $47\celsius$. The displacement rate decreases with increasing temperature due to the destabilization of the toehold. However, the probability of successful branch migration increases with higher temperatures. We therefore expect that for the shortest toeholds, the displacement will be accelerated with increasing temperature, consistent with experimental observations for DNA \cite{reynaldo2000kinetics}. 


The oxRNA model used for this work does not include explicit electrostatic interactions and was parametrized to RNA thermodynamics at $1\,\rm{M}$ salt.
It is possible to extend our model by including a Debye-Huckel potential parametrized to reproduce thermodynamics at lower salt concentrations \cite{wang2014modeling}. 
As the displacement involves two strands in close proximity during branch migration, it is expected that decreasing the salt concentration will have further effects on the rate of the displacement in addition to destabilizing the toehold.
This problem, along with the role of mismatches in the toehold region, will be the subject of further work.

Our model currently allows only Watson-Crick and wobble base pairs. It is possible that other kinds of tertiary structure contacts (such as ribose zippers or Hoogsteen base pairs)
could influence the branch migration process. We have also disallowed misbonds in our simulations, meaning that no mismatched bonds between the substrate and the 
invading strand and incumbent strand could occur. However, sequences used in the experiments are usually designed to avoid such misbonds, and hence we would not expect misbonds to affect the conclusions drawn from our simulations. It would be interesting in the future to study the effect of misbonds and secondary structure formation on the full toehold-mediated displacement reaction.

When designing an RNA displacement reaction for applications {\it in vivo}, the number of possible sequences is greatly reduced due to biological restrictions on the sequence that has to be recognized by the
interacting strands. Therefore, it is important to understand the factors that influence the kinetic rates of the underlying reaction. Our work hence provides estimates of how the rate of the reaction can be further enhanced, such as by placing the toehold at the $5^{\prime}$ end or increasing the toehold length.

The simulation code implementing the oxRNA model is freely available for download at dna.physics.ox.ac.uk as a part of the oxDNA software package.

We thank Erik Winfree, Niranjan Srinivas and Terence Hwa for useful discussions and Lorenzo Rovigatti for his contributions to the oxDNA simulation code development.
The authors acknowledge the Engineering and Physical
Sciences Research Council and University College (Oxford) for financial support.


\bibliographystyle{aip}

\bibliography{rna_biblio4}

 \setcounter{figure}{0}
 \makeatletter 
 \renewcommand{\thefigure}{S\@arabic\c@figure}
 \setcounter{equation}{0}
 \renewcommand{\theequation}{S\@arabic\c@equation}
 \setcounter{table}{0}
 \renewcommand{\thetable}{S-\@arabic\c@table}
  \setcounter{section}{0}
 \renewcommand{\thesection}{S-\@Roman\c@section}

\appendix 
\section*{Supplementary Material}

\subsection*{1D-model of strand displacement}
A simple 1-D model for the strand-displacement reaction was developed in Ref.~\onlinecite{srinivas2013biophysics} for DNA.  
For completeness, we repeat here the estimation of the displacement reaction rate with this simple model.

We assume that the rate at which the invading strand attaches to the toehold is $k_{\rm on}$, which is concentration-dependent. 
The invading strands then proceeds to successfully displace the incumbent strand during branch migration with a probability $p_{\rm bm | toe}$, so the rate of the successful displacement is 
\begin{equation}
\label{eq_kacko}
 k = k_{\rm on} \times p_{\rm bm | toe}. 
\end{equation}

The branch migration process is initiated by displacing the first base of the incumbent strand with a rate $k_{\rm first}$. The probability 
of initiating the branch migration rather than falling off is  
\begin{equation}
p_{\rm in} =  \frac{k_{\rm first} }{ k_{\rm first} + k_{\rm off} }, 
 \end{equation}
 where $k_{\rm off}$
is the rate for the invading strand to fall off once it is attached by the toehold. 

We assume that there are $b$ bases between the incumbent strand and the invading strand.
Once the branch migration is initiated and the invading strand has displaced the first base, it has a probability $1/(b-1)$ of successfully completing the displacement, where we assumed that the branch migration is a random walk where the invading strand can gain or lose a base pair with equal probability.  When the invading strand has displaced one base, the probability that the invading strand goes back to being bound just by the toehold is $1 - 1/(b-1)$. It can then again initiate displacement with probability $p_{\rm in}$.
We can hence approximate $p_{\rm bm | toe}$ as 
\begin{equation} 
p_{\rm bm | toe} = \frac{k_{\rm first} }{ k_{\rm first} + k_{\rm off} } \left(  \frac{1}{b-1}   + \frac{b-2}{b-1} \times p_{\rm bm | toe}  \right) ,
\end{equation}
from which we obtain
\begin{equation}
\label{eq_pbm}
 p_{\rm bm | toe} = \frac{k_{\rm first}}{k_{\rm first} + (b-1) k_{\rm off}} 
\end{equation}

From Eqs.~\ref{eq_pbm} and \ref{eq_kacko} we hence obtain
\begin{equation}
\label{eq_k}
  k = k_{\rm on} \times \frac{k_{\rm first}}{k_{\rm first} + (b-1) k_{\rm off}} = \frac{k_{\rm on}}{1  + (b-1) \frac{k_{\rm off}}{k_{\rm first}}}.
\end{equation}
In the limit of short toeholds, before the saturation regime of the displacement reaction is reached, we can assume $1 \ll (b-1) \frac{k_{\rm off}}{k_{\rm first}}$ and simplify Eq.~\ref{eq_k} to
\begin{equation}
 k \approx \frac{k_{\rm on}}{k_{\rm off}} \times \frac{k_{\rm first}}{b-1},
\end{equation}
which is used in the main text, where we used $k_{\rm bm} = \frac{k_{\rm first}}{b-1}$.

For long toeholds, when the saturation regime is reached, we can assume $1 \gg (b-1) \frac{k_{\rm off}}{k_{\rm first}}$ and the displacement rate becomes
\begin{equation}
 k \approx k_{\rm on}.
\end{equation}

\subsection*{FFS Simulations}
The Forward Flux Sampling (FFS) algorithm \cite{Allen2009} allows for efficient simulations of a transition from an initial (meta)stable state A (denoted as a state $Q = -2$ for our system) to a final stable state B ($Q=Q_{\rm max}$, where $Q_{\rm max}$ will be either 3 or 4 for our system, depending on the choice of order parameters as discussed in the following section).
We use FFS to calculate the fluxes from a state where the invading strand is unbound to a state where the invading strand successfully binds to the substrate and removes the incumbent strand. FFS introduces a series of interfaces $\lambda$ in the state space between the A and B states. First, a `brute-force' simulation is run to estimate the flux of trajectories that leave state A and cross the first interface. Then, one selects at random a crossing point at the first interface $\lambda^{0}_{-1}$ that was obtained from the generated trajectories and propagates it further until it either returns to the state A or reaches the second interface $\lambda_0^{1}$. By repeating this process one obtains an ensemble of points at the interface $\lambda_0^{1}$ and an estimate of the probability $P\left(\lambda_0^1 | \lambda_{-1}^0 \right)$ of reaching the interface $\lambda_0^1$. The procedure is repeated iteratively for the subsequent interfaces, and one can then estimate the transition rate from state A ($Q=-2$) to state B ($Q=Q_{\rm max}$) as  
\begin{equation}
 k_{AB} =  \phi_{-1}^0 \prod_{Q = 0}^{Q = Q_{\rm max}} P\left( \lambda_Q^{Q+1} | \lambda_{Q-1}^{Q} \right) 
\end{equation}
where $\phi_{-1}^0$ is the flux of trajectories leaving state $Q=-2$ and crossing the interface $\lambda^{0}_{-1}$. 

We previously used the FFS algorithm to study the kinetics of reactions involving DNA strands with the oxDNA model, and detailed description of this approach can be found in Refs.~\onlinecite{srinivas2013biophysics}, \onlinecite{ouldridge2013dna} and \onlinecite{schreck2014dna}.

\subsubsection*{Order parameters}

\begin{table}
\centering
\begin{tabular}{ c c }
  \hline  \hline   Q & Definition \\
  \hline
   -2 & $d > 3.36$\\
   -1 & $0.84 < d \leq 3.36$  \\
    0 & $d \leq 0.84$ \& $b = 0$ \\
   1 & $b \geq 1$  \\
   2 & $ b \geq l $\\
   3 & $ b \geq l+3$ \\
   4 & $b = 10 + l$ \& $d_2 > 3.36$ \\ \hline \hline 
\end{tabular}
\caption{The definition of order parameters Q for simulations with toehold lengths $l = 3$ or smaller. $d$ is the minimum distance between complementary bases in the invading strand and the toehold on the substrate and $b$ is the number of bonds between the invading strand and the substrate. $d_2$ is the minimum distance between all complementary bases of the incumbent strand and the substrate. All distances are in nm. Substrate is $10 + l$ bases long and toehold is $l$ base long. For toehold of length $l=1$, the order parameter $Q=2$ is defined as $b \geq 2$.}
\label{tab_ops}
\end{table}

\begin{table}
\centering
\begin{tabular}{ c c }
  \hline  \hline   Q & Definition \\
  \hline
   -2 & $d > 3.36$\\
   -1 & $0.84 < d \leq 3.36$  \\
   0 & $d \leq 0.84$ \& $b = 0$ \\
   1 & $b \geq 1$  \\
   2 & $ b \geq l $\\
   3 & $b = 10 + l$ \& $d_2 > 3.36$ \\ \hline \hline 
\end{tabular}
\caption{The definition of order parameters Q for simulations with toehold lengths $l$. For toehold of length $l=1$, the order parameter $Q=2$ is defined as $b \geq 2$. The definitions of variables in the table are the same as the ones used in Table~\ref{tab_ops}}
\label{tab_ops2}
\end{table}

The definition of the order parameters $Q$ used in the FFS simulations is provided in Tables~\ref{tab_ops} and \ref{tab_ops2}. We originally used order parameters from Table \ref{tab_ops2}, but due to limited accuracy we then decided to run simulations for 3-nucleotide (or smaller) toeholds with order parameters as defined in Table \ref{tab_ops}. The systems with order parameters from Table~\ref{tab_ops} have one additional interface compared to Table~\ref{tab_ops2}. The data obtained was consistent, and is included in Fig.~5 in the main text. 
If simulations using order parameters both from Table~\ref{tab_ops} and Table~\ref{tab_ops2} were considered for a given system, the reported FFS results are shown in two separate tables (one for each respective choice of order parameters). 
%

%

\subsubsection*{FFS simulation results}
The results of the forward flux sampling simulations are provided in Tables~\ref{tab_ops_6_3} to \ref{tab_ops_1_3} for toeholds of lengths 1 to 6 placed at either the $3^{\prime}$ or $5^{\prime}$ end at simulation temperature $37\celsius$. The results for simulations of a 3-base toehold at the $3^{\prime}$ end for temperatures $17\celsius$, $27\celsius$, and $47\celsius$ are shown in Tables~\ref{tab_ops_3_3_t17}, \ref{tab_ops_3_3_t27}, and \ref{tab_ops_3_3_t47}, respectively.
For each system considered, at least three independent simulations were carried out. The rates shown in Figs.~5 and 6 in the main text were obtained as the average of the rates from the simulations, with errorbars showing the maximum and minimum rates encountered. 

Tables \ref{tab_ops_6_3} to \ref{tab_ops_3_3_t47} show the number of successful crossings of a given interface along with the number of trajectories launched from the previous interface, obtained by summing all trajectories from all independent simulations for a given system. The probability of successful interface crossing shown in the tables is obtained as the number of crossings divided by the number of attempts. The numbers in brackets are the maximum absolute difference of the probability (or flux) shown in the table and the respective probabilities estimated from the individual independent simulations. For each interface of the individual simulations, the sampling was run until a desired number of successful crossings was reached.  Simulations from the same ensemble have the same number of successful crossings for each interface, which can be obtained as the number of crossings shown in the table divided by the number of independent simulations (up to small differences of few extra crossings for some simulations in tables~\ref{tab_ops_6_3}, \ref{tab_ops_6_5}, \ref{tab_ops_5_3}, and \ref{tab_ops_4_3} which resulted from the implementation details for a parallel architecture).

The description of the MD algorithm used for the sampling is provided in the methods section of the main text.

\begin{table}
\centering
\begin{tabular}{c c c l}
  \hline  \hline   $\lambda$ &  Crossings & Flux &   \\
  \hline
    $\lambda_{-1}^{0}$  & 9000  & \multicolumn{2}{c}{$\left( 1.39 \,(0.01) \right) \times 10^{-7}$ }\\ \hline  
   $\lambda$ & Success & Attempts & Probability \\ \hline 
   $\lambda^{1}_0$ & 9000 & 1372515 & $0.0066 \,(0.0007)$ \\
   $\lambda^{2}_1$ & 9022 & 71953 & $0.125 \,(0.04) $  \\
   $\lambda^{3}_2$ & 922 &  930 & $ 0.991 \,(0.003) $  \\  \hline \hline
\end{tabular}
\caption{FFS results for a 6-nucleotide toehold at the $3^{\prime}$ end. The data were obtained from 3 independent sets of simulations.}
\label{tab_ops_6_3}
\end{table}

\begin{table}
\centering
\begin{tabular}{c c c l}
  \hline  \hline   $\lambda$ &  Crossings & Flux &   \\
  \hline
    $\lambda_{-1}^{0}$  & 9000  & \multicolumn{2}{c}{$\left( 1.40 \,(0.03) \right) \times 10^{-7}$ }\\ \hline  
   $\lambda$ & Success & Attempts & Probability \\ \hline 
   $\lambda^{1}_0$ & 9000 & 1359767 & $0.0066 \,(0.0004)$ \\
   $\lambda^{2}_1$ & 9022 & 78126 & $0.115 \,(0.01) $  \\
   $\lambda^{3}_2$ & 900 &  907 & $ 0.992 \,(0.008) $  \\  \hline \hline
\end{tabular}
\caption{FFS results for a 6-nucleotide toehold at the $5^{\prime}$ end. The data were obtained from 3 independent sets of simulations.}
\label{tab_ops_6_5}
\end{table}

\begin{table}
\centering
\begin{tabular}{c c c l}
  \hline  \hline   $\lambda$ &  Crossings & Flux &   \\
  \hline
    $\lambda_{-1}^{0}$  & 9000  & \multicolumn{2}{c}{$\left( 1.28 \,(0.03)\right) \times 10^{-7}$ }\\ \hline  
   $\lambda$ & Success & Attempts & Probability \\ \hline 
   $\lambda^{1}_0$ & 9000 & 1520553 & $ 0.006 \,(0.001)$ \\
   $\lambda^{2}_1$ & 9015 & 115774 & $0.0779 \,(0.004)$  \\
   $\lambda^{3}_2$ & 918 &  1022 & $ 0.898 \,(0.003) $  \\  \hline \hline
\end{tabular}
\caption{FFS results for a 5-nucleotide toehold at the $3^{\prime}$ end. The data were obtained from 3 independent sets of simulations.}
\label{tab_ops_5_3}
\end{table}

\begin{table}
\centering
\begin{tabular}{c c c l}
  \hline  \hline   $\lambda$ &  Crossings & Flux &   \\
  \hline
    $\lambda_{-1}^{0}$  & 9000  & \multicolumn{2}{c}{$\left( 1.30 \,(0.02)\right) \times 10^{-7}$ }\\ \hline  
   $\lambda$ & Success & Attempts & Probability \\ \hline 
   $\lambda^{1}_0$ & 9000 & 1363822 & $ 0.0066 \,(0.0006)$ \\
   $\lambda^{2}_1$ & 9000 & 92789 & $0.1 \,(0.05)$  \\
   $\lambda^{3}_2$ & 900 &  946 & $ 0.951 \,(0.008) $  \\  \hline \hline
\end{tabular}
\caption{FFS results for a 5-nucleotide toehold at the $5^{\prime}$ end. The data were obtained from 3 independent sets of simulations.}
\label{tab_ops_5_5}
\end{table}

\begin{table}
\centering
\begin{tabular}{c c c l}
  \hline  \hline   $\lambda$ &  Crossings & Flux &   \\
  \hline
    $\lambda_{-1}^{0}$  & 9000  & \multicolumn{2}{c}{$\left( 1.53 \,(0.01)\right) \times 10^{-7}$ }\\ \hline  
   $\lambda$ & Success & Attempts & Probability \\ \hline 
   $\lambda^{1}_0$ & 12000 & 2777034 & $ 0.0043 \,(0.0007)$ \\
   $\lambda^{2}_1$ & 12001 & 128781 & $0.09 \,(0.04)  $  \\
   $\lambda^{3}_2$ &  903 & 2961  & $ 0.30 \,(0.01) $  \\  \hline \hline
\end{tabular}
\caption{FFS results for a 4-nucleotide toehold at the $3^{\prime}$ end. The data were obtained from 3 independent sets of simulations.}
\label{tab_ops_4_3}
\end{table}

\begin{table}
\centering
\begin{tabular}{c c c l}
  \hline  \hline   $\lambda$ &  Crossings & Flux &   \\
  \hline
    $\lambda_{-1}^{0}$  & 9000  & \multicolumn{2}{c}{$\left( 1.16 \,(0.02)\right) \times 10^{-7}$ }\\ \hline  
   $\lambda$ & Success & Attempts & Probability \\ \hline 
   $\lambda^{1}_0$ & 12000 & 2140159 & $ 0.0056 \,(0.0004)$ \\
   $\lambda^{2}_1$ & 12000 & 80545 & $0.15 \,(0.04)  $  \\
   $\lambda^{3}_2$ &  900 & 1961  & $ 0.46 \,(0.03) $  \\  \hline \hline
\end{tabular}
\caption{FFS results for a 4-nucleotide toehold at the $5^{\prime}$ end. The data were obtained from 3 independent sets of simulations.}
\label{tab_ops_4_5}
\end{table}

\begin{table}
\centering
\begin{tabular}{c c c l}
  \hline  \hline   $\lambda$ &  Crossings & Flux &   \\
  \hline
    $\lambda_{-1}^{0}$  & 9000  & \multicolumn{2}{c}{$\left( 1.05 \,(0.04) \right) \times 10^{-7}$ }\\ \hline  
   $\lambda$ & Success & Attempts & Probability \\ \hline 
   $\lambda^{1}_0$ & 18000 & 3432942 & $ 0.005 \,(0.0005)$ \\
   $\lambda^{2}_1$ & 18000 & 288833 & $0.062 \,(0.004)  $  \\
   $\lambda^{3}_2$ &  900 & 25303  & $ 0.036 \,(0.004) $  \\  \hline \hline
\end{tabular}
\caption{FFS results for a 3-nucleotide toehold at the $5^{\prime}$ end. The data were obtained from 3 independent sets of simulations.}
\label{tab_ops_3_5a}
\end{table}

\begin{table}
\centering
\begin{tabular}{c c c l}
  \hline  \hline   $\lambda$ &  Crossings & Flux &   \\
  \hline
    $\lambda_{-1}^{0}$  & 1730  & \multicolumn{2}{c}{ $1.06  \times 10^{-7}$ }\\ \hline  
   $\lambda$ & Success & Attempts & Probability \\ \hline 
   $\lambda^{1}_0$ & 18000 & 3432942 & $ 0.0056 $ \\
   $\lambda^{2}_1$ & 18000 & 288833 & $0.088  $  \\
   $\lambda^{3}_2$ &  900 & 25303  & $ 0.34$ \\
   $\lambda^{4}_3$ &  900 & 25303  & $ 0.14$  \\  \hline \hline
\end{tabular}
\caption{FFS results for a 3-nucleotide toehold at the $5^{\prime}$ end. The data were obtained from 1 set of simulations.}
\label{tab_ops_3_5b}
\end{table}


\begin{table}
\centering
\begin{tabular}{c c c l}
  \hline  \hline   $\lambda$ &  Crossings & Flux &   \\
  \hline
    $\lambda_{-1}^{0}$  & 9000  & \multicolumn{2}{c}{$\left( 1.00 \,(0.01) \right) \times 10^{-7}$ }\\ \hline  
   $\lambda$ & Success & Attempts & Probability \\ \hline 
   $\lambda^{1}_0$ & 18000 & 3344529 & $ 0.0054 \,(0.0005)$ \\
   $\lambda^{2}_1$ & 18000 & 282819 & $0.06 \,(0.08)  $  \\
   $\lambda^{3}_2$ &  900 & 39767  & $ 0.02 \,(0.01) $  \\  \hline \hline
\end{tabular}
\caption{FFS results for a 3-nucleotide toehold at the $3^{\prime}$ end. The data were obtained from 3 independent sets of simulations.}
\label{tab_ops_3_3a}
\end{table}

\begin{table}
\centering
\begin{tabular}{c c c l}
  \hline  \hline   $\lambda$ &  Crossings & Flux &   \\
  \hline
    $\lambda_{-1}^{0}$  & 6000  & \multicolumn{2}{c}{ $\left(1.03 \,(0.01) \right) \times 10^{-7}$ }\\ \hline  
   $\lambda$ & Success & Attempts & Probability \\ \hline 
   $\lambda^{1}_0$ & 12000 & 2593563 & $ 0.0046 \,( 0.0002)  $ \\
   $\lambda^{2}_1$ & 12000 & 339889 & $0.035 \,(0.008)  $  \\
   $\lambda^{3}_2$ &  1000 &  9717  & $ 0.10 \,(0.01) $ \\
   $\lambda^{4}_3$ &  600 & 2577  & $ 0.233 \,(0.006)   $  \\  \hline \hline
\end{tabular}
\caption{FFS results for a 3-nucleotide toehold at the $3^{\prime}$ end. The data were obtained from 2 sets of simulations.}
\label{tab_ops_3_3b}
\end{table}


\begin{table}
\centering
\begin{tabular}{c c c l}
  \hline  \hline   $\lambda$ &  Crossings & Flux &   \\
  \hline
    $\lambda_{-1}^{0}$  & 6000  & \multicolumn{2}{c}{$\left( 0.9 \,(0.003)\right) \times 10^{-7}$ }\\ \hline  
   $\lambda$ & Success & Attempts & Probability \\ \hline 
   $\lambda^{1}_0$ & 12000 & 3304853 & $ 0.0036 \,(0.0004) $ \\
   $\lambda^{2}_1$ & 12000 & 64190  & $0.19 \,(0.02) $  \\
   $\lambda^{3}_2$ &  600 & 810861  & $ 0.0007 \,(0.003) $  \\  \hline \hline
\end{tabular}
\caption{FFS results for a 2-nucleotide toehold at the $5^{\prime}$ end. The data were obtained from 2 independent sets of simulations.}
\label{tab_ops_2_5a}
\end{table}

\begin{table}
\centering
\begin{tabular}{c c c l}
  \hline  \hline   $\lambda$ &  Crossings & Flux &   \\
  \hline
    $\lambda_{-1}^{0}$  & 9000  & \multicolumn{2}{c}{$\left( 0.87 \,(0.01)\right) \times 10^{-7}$ }\\ \hline  
   $\lambda$ & Success & Attempts & Probability \\ \hline 
   $\lambda^{1}_0$ & 18000 & 3657559 & $ 0.0049 \,(0.0004) $ \\
   $\lambda^{2}_1$ & 18000 & 130705  & $ 0.14 \,(0.19) $  \\
   $\lambda^{3}_2$ &  1500 & 88745  & $ 0.0169 \,(0.004) $  \\  
   $\lambda^{4}_3$ &  900 &  15585 & $ 0.0577 \,(0.009) $  \\  \hline \hline
\end{tabular}
\caption{FFS results for a 2-nucleotide toehold at the $5^{\prime}$ end. The data were obtained from 3 independent sets of simulations.}
\label{tab_ops_2_5b}
\end{table}

\begin{table}
\centering
\begin{tabular}{ c c c l}
  \hline  \hline   $\lambda$ &  Crossings & Flux &   \\
  \hline
    $\lambda_{-1}^{0}$  & 9000  & \multicolumn{2}{c}{$\left( 0.78 \,(0.02)\right) \times 10^{-7}$ }\\ \hline  
   $\lambda$ & Success & Attempts & Probability \\ \hline 
   $\lambda^{1}_0$ & 18000 & 4683647  & $ 0.0038 \,(0.0004) $ \\
   $\lambda^{2}_1$ & 18000 & 219009 & $ 0.08 \,(0.04) $  \\
   $\lambda^{3}_2$ &  1500 & 930597 & $ 0.0016 \,(0.0006) $  \\  
   $\lambda^{4}_3$ &  900 & 6234  & $ 0.144  \,(0.006) $  \\  \hline \hline
\end{tabular}
\caption{FFS results for a 2-nucleotide toehold at the $3^{\prime}$ end. The data were obtained from 3 independent sets of simulations.}
\label{tab_ops_2_3b}
\end{table}

\begin{table}
\centering
\begin{tabular}{ c c c l}
  \hline  \hline   $\lambda$ &  Crossings & Flux &   \\
  \hline
    $\lambda_{-1}^{0}$  & 6000  & \multicolumn{2}{c}{$\left( 0.79 \,(0.01)\right) \times 10^{-7}$ }\\ \hline  
   $\lambda$ & Success & Attempts & Probability \\ \hline 
   $\lambda^{1}_0$ & 12000 & 2780446  & $ 0.004 \,(0.001) $ \\
   $\lambda^{2}_1$ & 12000 & 426468 & $ 0.03 \,(0.09) $  \\
   $\lambda^{3}_2$ &  600 &  1421539 & $ 0.0004  \,(0.0001) $  \\  \hline \hline
\end{tabular}
\caption{FFS results for a 2-nucleotide toehold at the $3^{\prime}$ end. The data were obtained from 2 independent sets of simulations.}
\label{tab_ops_2_3a}
\end{table}

\begin{table}
\centering
\begin{tabular}{ c c c l}
  \hline  \hline   $\lambda$ &  Crossings & Flux &   \\
  \hline
    $\lambda_{-1}^{0}$  & 9000  & \multicolumn{2}{c}{$\left( 0.66 \,(0.07)\right) \times 10^{-7}$ }\\ \hline  
   $\lambda$ & Success & Attempts & Probability \\ \hline 
   $\lambda^{1}_0$ & 18000 & 4727390  & $ 0.0038 \,(0.0005) $ \\
   $\lambda^{2}_1$ & 18000 &  426011 & $  0.042 \,(0.015) $  \\
   $\lambda^{3}_2$ &  1500 & 579917 & $ 0.003 \,(0.03) $  \\  
   $\lambda^{4}_3$ &  900 & 132335  & $0.007 \,(0.009) $  \\  \hline \hline
\end{tabular}
\caption{FFS results for a 1-nucleotide toehold at the $5^{\prime}$ end. The data were obtained from 3 independent sets of simulations.}
\label{tab_ops_1_5b}
\end{table}

\begin{table}
\centering
\begin{tabular}{ c c c l}
  \hline  \hline   $\lambda$ &  Crossings & Flux &   \\
  \hline
    $\lambda_{-1}^{0}$  & 9000  & \multicolumn{2}{c}{$\left( 0.66 \,(0.09)\right) \times 10^{-7}$ }\\ \hline  
   $\lambda$ & Success & Attempts & Probability \\ \hline 
   $\lambda^{1}_0$ & 18000 & 5017739  & $ 0.0036 \,(0.0005) $ \\
   $\lambda^{2}_1$ & 18000 & 676294 & $  0.03 \,(0.04) $  \\
   $\lambda^{3}_2$ &  900 & 30649274 & $ 0.00003  \,(0.00003) $  \\  \hline \hline
\end{tabular}
\caption{FFS results for a 1-nucleotide toehold at the $3^{\prime}$ end. The data were obtained from 3 independent sets of simulations.}
\label{tab_ops_1_5a}
\end{table}


\begin{table}
\centering
\begin{tabular}{ c c c l}
  \hline  \hline   $\lambda$ &  Crossings & Flux &   \\
  \hline
    $\lambda_{-1}^{0}$  & 9000  & \multicolumn{2}{c}{$\left( 0.50 \,(0.01)\right) \times 10^{-7}$ }\\ \hline  
   $\lambda$ & Success & Attempts & Probability \\ \hline 
   $\lambda^{1}_0$ & 18000 & 6935564 & $ 0.0026 \,(0.0002) $ \\
   $\lambda^{2}_1$ & 18000 &  2165417 & $ 0.01 \,(0.02) $  \\
   $\lambda^{3}_2$ &  1500 &  2319850 & $ 0.00065 \,(0.00002) $  \\  
   $\lambda^{4}_3$ &  900 &  18734 & $ 0.05  \,(0.03) $  \\  \hline \hline
\end{tabular}
\caption{FFS results for a 1-nucleotide toehold at the $3^{\prime}$ end. The data were obtained from 3 independent sets of simulations.}
\label{tab_ops_1_3}
\end{table}


\begin{table}
\centering
\begin{tabular}{c c c l}
  \hline  \hline   $\lambda$ &  Crossings & Flux &   \\
  \hline
    $\lambda_{-1}^{0}$  & 9000  & \multicolumn{2}{c}{ $\left(0.77 \,(0.02) \right) \times 10^{-7}$ }\\ \hline  
   $\lambda$ & Success & Attempts & Probability \\ \hline 
   $\lambda^{1}_0$ & 18000 & 4119731 & $ 0.004 \,( 0.002)  $ \\
   $\lambda^{2}_1$ & 1800 & 251974 & $0.07 \,(0.02)  $  \\
   $\lambda^{3}_2$ &  1500 &  5547  & $ 0.27 \,(0.06) $ \\
   $\lambda^{4}_3$ &  900 & 3657  & $ 0.25 \,(0.01)   $  \\  \hline \hline
\end{tabular}
\caption{FFS results for a 3-nucleotide toehold at the $3^{\prime}$ end. The data were obtained from 3 independent sets of simulations at $17\celsius$.}
\label{tab_ops_3_3_t17}
\end{table}

\begin{table}
\centering
\begin{tabular}{c c c l}
  \hline  \hline   $\lambda$ &  Crossings & Flux &   \\
  \hline
    $\lambda_{-1}^{0}$  & 9000  & \multicolumn{2}{c}{ $\left(0.87 \,(0.01) \right) \times 10^{-7}$ }\\ \hline  
   $\lambda$ & Success & Attempts & Probability \\ \hline 
   $\lambda^{1}_0$ & 18000 & 3843169 & $ 0.005\,( 0.001)  $ \\
   $\lambda^{2}_1$ & 1800 & 251058 & $0.07 \,(0.02)  $  \\
   $\lambda^{3}_2$ &  1500 &  9490  & $ 0.16 \,(0.01) $ \\
   $\lambda^{4}_3$ &  900 & 3795  & $ 0.24 \,(0.01)   $  \\  \hline \hline
\end{tabular}
\caption{FFS results for a 3-nucleotide toehold at the $3^{\prime}$ end. The data were obtained from 3 independent sets of simulations at $27\celsius$.}
\label{tab_ops_3_3_t27}
\end{table}

\begin{table}
\centering
\begin{tabular}{c c c l}
  \hline  \hline   $\lambda$ &  Crossings & Flux &   \\
  \hline
    $\lambda_{-1}^{0}$  & 9000  & \multicolumn{2}{c}{ $\left(1.10 \,(0.03) \right) \times 10^{-7}$ }\\ \hline  
   $\lambda$ & Success & Attempts & Probability \\ \hline 
   $\lambda^{1}_0$ & 18000 & 3766518 & $ 0.005\,( 0.001)  $ \\
   $\lambda^{2}_1$ & 1800 & 634145 & $0.028 \,(0.01)  $  \\
   $\lambda^{3}_2$ &  1500 &  19227  & $ 0.078 \,(0.03) $ \\
   $\lambda^{4}_3$ &  900 & 3599  & $ 0.25\,(0.01)   $  \\  \hline \hline
\end{tabular}
\caption{FFS results for a 3-nucleotide toehold at the $3^{\prime}$ end. The data were obtained from 3 independent sets of simulations at $47\celsius$.}
\label{tab_ops_3_3_t47}
\end{table}

\clearpage

\end{document}